\documentstyle [preprint,revtex]{aps}

\begin{document}

\hfill {MIT-CTP-2108}

\hfill {ORNL-CCIP-92-06}

\hfill {UTK-92-4}

\hfill {UMS/HEP 92-021}

\begin{title}
I=3/2 $K \pi $ Scattering in the\\
Nonrelativistic Quark Potential Model\\
\end{title}

\author{T. Barnes}

\begin{instit}
Physics Division and Center for Computationally Intensive Physics\\
Oak Ridge National Laboratory, Oak Ridge, TN 37831-6373\\
and\\
Department of Physics\\
University of Tennessee, Knoxville, TN 37996-1200\\
\end{instit}

\author{E.S. Swanson}

\begin{instit}
Center for Theoretical Physics\\
Laboratory of Nuclear Science and Department of Physics\\
Massachusetts Institute of Technology, Cambridge, MA 02139\\
\end{instit}

\author {J. Weinstein}

\begin{instit}
Department of Physics and Astronomy\\
University of Mississippi\\
University, MS 38677\\
\end{instit}

\begin{abstract}
We study $I=3/2$ elastic $K\pi $  scattering to Born order
using nonrelativistic quark wavefunctions in a constituent-exchange model.
This channel is ideal for the study of nonresonant meson-meson scattering
amplitudes
since s-channel resonances do not contribute significantly.
Standard
quark model parameters yield good agreement with the measured S- and P-wave
phase shifts and with PCAC calculations of the scattering length.
The P-wave phase shift is especially interesting because it is nonzero
solely due to
$SU(3)_f$ symmetry breaking effects, and is found to be in good agreement
with experiment given conventional values for the strange and nonstrange
constituent quark masses.
\end{abstract}

\centerline{Submitted to {\sl Phys. Rev.} {\bf D}}
\centerline{June 1992}
\newpage

\section{Introduction}

The derivation of hadronic interactions from QCD has been a goal of nuclear
physics for many years. At present this appears to be a very difficult problem;
even the
more modest goal of deriving hadronic interactions from the nonrelativistic
quark model is difficult, due in part
to the variety of mechanisms which contribute to
scattering.
In a typical hadronic scattering process these mechanisms include
s-channel resonance production, t-channel
resonance exchange, and nonresonant scattering.
Despite the apparent complexity, there is considerable evidence
that some scattering amplitudes
are dominated by relatively simple
perturbative QCD processes.
One well known example is the
short-range part of the
NN interaction; many
groups have concluded that the NN repulsive core is due to the combined
effects of
the Pauli principle and the color magnetic spin-spin component of one gluon
exchange.
Similarly, the
intermediate-range
attractive interaction may be due to a relatively simple effect at the
quark level, specifically a color-dipole interaction induced by the spatial
distortion of the three-quark clusters \cite{NIcart}.
Of course one pion exchange dominates at sufficiently large
separations, and in a more complete description one should adjoin this to
the short-range quark interaction.

In this paper we discuss $I = 3/2$ $K\pi $ elastic scattering in the
nonrelativistic quark model. This process resembles $NN$
scattering in that s-channel resonances are not expected to give important
contributions, assuming that multiquark resonances are not in evidence.
The Born-order QCD scattering amplitude for this process involves
one gluon exchange followed by quark exchange.
In a previous paper \cite{BS} it was shown that this simple
description of hadron-hadron scattering leads to
good agreement with the nonperturbative variational results
of Weinstein and Isgur
\cite{WI} near threshold,
and with the measured $I=2$ $\pi\pi$ S-wave phase shift throughout the
full range of $M_{\pi\pi}$ for which data exists. A
similar method was used in Ref.\ \cite{Swan}
to extract effective potentials for many
low-lying
meson-meson channels.  These
potentials have recently been applied to several
problems in low-energy meson
physics. In particular, Dooley {\it et al.} \cite{DSB} used the results
of Ref.\ \cite{Swan} to suggest that the $IJ^{PC} = 0 0^{++}$ $\theta
(1720)$ may be a  $(K^*\bar K^*)$-$(\omega\phi)$ vector-vector molecular
bound state. Simple estimates of
branching ratios in this model find
good agreement with Particle Data Group values
and predict new decay modes. In another application it has been
argued that the $f_1(1420)$ ``E" effect may be a threshold enhancement which is
due to an attractive $(K^* \bar K)$-$(\omega \phi)$ interaction
in the $01^{++}$ channel \cite{Swan}.

The $I= 3/2$ $K\pi $ channel is an ideal one for testing our model
of Born scattering amplitudes,
since we expect it to be largely unaffected by s-channel resonances.
The success of the
previous application of the model to $I = 2$ $\pi\pi$ scattering
and the assumption of flavor symmetry suggest
that we may also find reasonable agreement with the experimental
S-wave $K\pi $ phase shift. The P-wave, however, is driven entirely by
flavor symmetry {\it breaking}, and hence the interplay of these two waves
provides an interesting and nontrivial test of the model.

\section{Method}

The calculation is based on a standard quark model Hamiltonian of the
form

\begin{equation}
H = \sum_{i=1}^4 \left( {p_i^2 \over 2 m_i} + m_i \right) + \sum_{i<j}
\left[ V_{conf}(r_{ij}) + V_{cont}(r_{ij})\; \vec S_i \cdot \vec S_j +
V_{orb} \; \right] \vec F_i \cdot \vec F_j
\end {equation}
where
\begin{equation}
V_{cont} = {-8 \pi \alpha_s\over 3 m_i m_j}\; \delta(\vec r_{ij})
\end {equation}
is the contact color-hyperfine interaction, $V_{conf}$ is a
confinement potential, and $V_{orb}$ represents spin-orbit and tensor
interactions.
We
neglect $V_{orb}$ in this paper since its effects are generally found to
be numerically small
in meson spectroscopy \cite{GI}, and are expected to be unimportant in
the scattering of two $\ell=0$ mesons as well.
We shall also ignore the contribution of $V_{conf}$ to the scattering
interaction.
This may appear to be a questionable approximation; however,
resonating group calculations have found that the exchange (scattering)
kernel due to $V_{conf}$ is much smaller than the corresponding
kernel for the
hyperfine term \cite{Shim}. This result was also found in the
variational
calculation of Ref.\ \cite{WI} and the perturbative calculation of Ref.\
\cite{Swan}. The latter reference noted that the small $V_{conf}$
contribution to scattering is due to a color-factor
cancellation in the matrix element of $V_{conf}$.
However, this result only applies to certain channels;
one should {\it not} neglect the
effects of the confinement term in scattering involving
vector mesons.

Although we calculate the scattering amplitude only
to Born order, there is
evidence that this is a useful and even accurate approximation
in systems which are not dominated by s-channel resonances or t-channel
meson exchange.
First, as there is
little evidence for flavor mixing
in meson spectroscopy outside the $\eta-\eta'$ system,
one anticipates that neglecting higher terms in
the Born series (such as $q \bar q \rightarrow gg \rightarrow q \bar q$)
is not a bad approximation. In addition, the Born-approximation
$I=2$ $\pi\pi$ effective
potentials
derived in Refs.\ \cite{BS} and \cite{Swan} are numerically very similar
to the
nonperturbative potentials derived by Weinstein and Isgur.
Finally, comparison of perturbative phase shifts to those found in a
variational resonating group calculation shows good numerical
agreement \cite{Swan}.

For simplicity we use single Gaussians for the asymptotic pion and kaon
wavefunctions,
\begin{equation}
\psi_{\pi(K)}(r) = \left({\beta_{\pi(K)}^2 \over \pi}\right)^{3/4}
{\rm e}^{-\beta_{\pi(K)}^2\, r^2/2} \ ,
\end {equation}
where $r = |\vec r_q - \vec r_{\bar q}|$.
The corresponding momentum-space wavefunction $\phi(k_{rel})$ is
a function of the magnitude of the relative momentum vector
$\vec k_{rel}=(m_{\bar q}\vec k_q - m_q\vec k_{\bar q})/(m_q + m_{\bar q})$.

Flavor symmetry breaking is incorporated through unequal strange
and nonstrange quark masses (we introduce a mass ratio
$\rho = m_u/m_s$) and a meson width parameter $\xi$
(defined by $\xi = \beta_\pi^2/\beta_K^2$;
$\xi < 1$ corresponds to a smaller kaon than
pion).
Of course these parameters are related.
For instance if
we take $V_{conf}(r_{ij}) = C + \kappa r_{ij}^2 /2$  then
$\xi = \sqrt{(1 + \rho)/2}$.
Standard quark model values for the constituent masses,
$m_u = 0.33$ GeV and $m_s =
0.55$ GeV, give $\rho = 0.6$, and from the SHO relation above
we might anticipate
$\xi \approx 0.9$.
A fit to light meson spectroscopy in a
Coulomb plus linear potential model with a contact hyperfine term
finds a similar $\rho $ value of $\rho = 0.58$.  With this $\rho$
a single-Gaussian variational calculation
\cite{Swan} actually finds a value for $\xi$ slightly above unity,
$\xi = 1.05$, because the stronger pion hyperfine attraction leads to a smaller
pion than kaon despite the heavier strange quark mass. In any case we expect
$\xi$ to be near unity.

There are four Born-order quark exchange graphs for $K\pi $ scattering,
which
we previously classified as two
``transfer"  or ``capture" processes in our discussion of $\pi\pi$
scattering \cite{BS}.
The transfer diagrams represent scattering due to a
spin-spin hyperfine interaction between a
quark pair ($T_1$) or an antiquark pair ($T_2$).
In the capture
diagrams the interaction is between a quark-antiquark pair in different mesons,
$u\bar s$ for $C_1$ and $u \bar d$ for $C_2$. We apply the
methods of Ref. \cite{BS} (Appendix C)
to obtain the Born-order Hamiltonian matrix element
$h_{fi}$ for these diagrams, which is

\begin{equation}
h_{fi} = {1\over (2\pi)^3}
{4 \pi \alpha_s \over 9 m_u^2} \Big( T_1 + T_2 + C_1 + C_2 \Big) \ ,
\end {equation}
where the term contributed by each diagram is

\begin{mathletters}
\begin{eqnarray}
T_1 &=& \exp\bigg\{
- (1 + \xi(1+\zeta)^2 )\;
\bigg[ {1- \mu\over 2} \bigg] \;
{k^2 \over 4 \beta_\pi^2}\;
\bigg\} \\
T_2 &=& \rho \left({2\sqrt{\xi} \over 1 + \xi } \right)^{3}
\exp\bigg\{ -
{\xi \over 1+\xi} \; \bigg[ 1 + (1-\zeta)^2 + 2(1-\zeta)\mu \, \bigg] \;
{k^2 \over 4\beta_\pi^2} \; \bigg\} \\
C_1 &=& \rho \left({4 \over 2 + \xi}\right)^{3/2} \exp \bigg\{
-{ 1 \over 2+\xi} \; \bigg[ 1+ 3\xi - \xi \zeta (1-\zeta)
+ ( \xi - 1- 3 \xi \zeta) \mu \,
\bigg]  \; {k^2 \over 4\beta_\pi^2}\; \bigg\} \\
C_2 &=& \left({4\xi \over 1 +2\xi}\right)^{3/2} \exp \bigg\{
-{ \xi \over 1+2\xi} \;
\bigg[ 3 - \zeta + \zeta^2 + \xi (1+\zeta)^2 \nonumber \\
&& + (1 - 3\zeta -\xi (1+\zeta)^2 ) \mu \, \bigg]
{k^2 \over 4\beta_\pi^2} \; \bigg\} \ .\\
\nonumber
\end{eqnarray}
\end{mathletters}
In these matrix elements $\mu = \cos(\theta_{CM})$, where $\theta_{CM}$ is
the center of mass
scattering angle, the quark mass parameter $\zeta$ is $
(m_s-m_u)/(m_s+m_u) = (1-\rho)/(1+\rho)$, and $k$ is the magnitude of
the asymptotic three-momentum of each meson in the CM frame.
The matrix element $h_{fi}$ is related to the $\ell$th partial-wave
phase shift by \cite{BS}
\begin{equation}
\delta^{(\ell)} = {- 2\pi^2 k E_\pi E_K \over ( E_\pi + E_K)} \int_{-1}^{1}
h_{fi}(\mu)\, P_\ell (\mu)\,d\mu \ ,
\end{equation}
with the meson energies related to $k$ by relativistic kinematics.
This result involves $\delta^{(\ell )} $ rather than
$\sin \delta^{(\ell)} $ because we choose to
equate our Born amplitude to the leading
term in the
elastic scattering amplitude $(\exp\{2i\delta^{(\ell)}\}-1)/2i$
rather than to the full real part. The phase shifts for all partial waves
follow
from this result
through application of the integral $\int_{-1}^1 e^{a\mu} P_\ell (\mu)
d\mu = 2 i_\ell (a)$ where $i_\ell$ is the modified spherical Bessel function
of the first kind.

\section{Results and Discussion}

On evaluating the angular integrals (6) we find
$I=3/2$ $K\pi $ phase shifts for all $\ell$ in Born
approximation given SHO wavefunctions; these are functions of the
four free parameters
$\beta_\pi$, $\alpha_s/m_u^2$, $\rho=m_u/m_s$, and $\xi=\beta_\pi^2/\beta_K^2$,
and require the physical meson masses as input.
In the following discussion we shall fix the nonstrange
quark mass to be
$m_u=0.33$ GeV since the phase shifts actually involve the ratios given
above rather than the absolute scale of $m_u$.

We proceed by fitting the predicted phase shifts to the S- and P-wave phase
shift data of Estabrooks {\it et al.} \cite{Esta}.
Note however that there may be a discrepancy between this
data and earlier $I=3/2$ $K\pi$ results \cite{Jong,LP} near threshold;
two S-wave data sets are shown in Fig.\ \ref{fig1}.

As an initial ``benchmark" prediction we first
neglect flavor-symmetry violation (except for the use of
physical meson masses in kinematics and phase space)
and employ the same
parameters we previously used to describe
$I=2$ $\pi\pi$ scattering in Ref.\ \cite{BS};
$\alpha_s = 0.6$, $m_u = 0.33$ GeV and $\beta_\pi($fitted to $\pi\pi)
= 0.337$ GeV,
and we
set $m_s=m_u$ and $\beta_K=\beta_{\pi}$ so that $\rho = 1$ and $\xi = 1$.
The resulting S-wave phase shift is shown as
a dotted line in Fig.\ \ref{fig1}. Although the shape of the predicted phase
shift is qualitatively correct, evidently the predicted magnitude is
somewhat larger
than the data at invariant masses above 0.9 GeV.

Of course this flavor-symmetric
parameter set is unrealistic because it does not assume a
heavier strange quark; allowing $m_s$ to vary
yields the value $\rho = 0.677$ in a fit to the S-wave data
of Estabrooks {\it et al.}; this is
close to the $\rho \approx 0.33 \ \hbox{GeV} / 0.55 \ \hbox{GeV}$ $= 0.6$
expected
from $q\bar q$ quark model spectroscopy. The resulting phase shift is shown as
a dashed line in Fig.\ \ref{fig1}, and the agreement is impressive. The same
parameter set gives a P-wave phase shift which is shown as a dashed line in
Fig.\ \ref{fig2}. Evidently the agreement with experiment
is again quite good. Note that
the predicted P-wave phase shift is zero for $\rho = 1$, so the S-wave data
are consistent with approximate flavor symmetry (which implies
$\pi\pi$ S-wave $\approx$ $K\pi $ S-wave) and the P-wave data are consistent
with the expected amount of
flavor symmetry breaking (seen in the nonzero $K\pi $ P-wave).

Although we have found a satisfactory description of the data simply by using
$\pi\pi$ parameters and physical meson masses
and fitting $m_s$, it is of interest to investigate the
sensitivity of our results to changes in the other parameters and to
determine their global optimum values.
Fixing $\xi = 1$ and $\beta_\pi = 0.337$ GeV and fitting $\alpha_s$ and
$\rho$ to the Estabrooks {\it et al.} S-wave data
gives $\alpha_s = 0.634$ and $\rho = 0.604$,
again consistent with standard quark model values.
A global fit to the 33 S- and P-wave data points of Estabrooks
{\it et al.}
with all four parameters free
gives $\rho = 0.789$, $\alpha_s = 0.577$, $\beta_\pi = 0.293$
GeV and $\xi = 0.568$. The rather large $m_u/m_s$ in this fit
is partially compensated by a spatially small kaon wavefunction, but as the
phase shifts are rather insensitive to $\xi$, and we expect a value
closer to unity, this best fit probably gives less realistic
parameter values than the
single-parameter fit which found $\rho=0.677$.
The phase shifts
predicted by the
the global four-parameter set are shown as
solid lines
in Figs.\
\ref{fig1} and \ref{fig2}.
Note that the four-parameter
S-wave phase shift is essentially indistinguishable
from the one-parameter $(m_s)$ fit (dashed line);
the most important difference in the predictions of the two parameter sets
is in the P-wave, which is not yet very well determined
experimentally.

Estabrooks {\it et al.} also reported measurements of the $I=3/2$ $K\pi $
D-wave phase shift. We predict a
small negative D-wave phase shift in accord with the data, although
the magnitude of our D-wave is somewhat smaller than is observed.
A similar discrepancy in the $I=2$ $\pi\pi$ D-wave was previously noted
\ \cite{BS,Swan}.
It should be stressed that the
D-wave is qualitatively different from the P-wave;
it is not driven by flavor symmetry breaking and is an intrinsically
small effect at these energies, so that other contributions which we have
neglected may be important here. Possible contributions to this higher
partial wave include
the confinement, spin-orbit and tensor interactions.
The departure of the actual $q\bar q$ wavefunction from the assumed single
Gaussian may also be important, although
this effect was investigated in Ref.\ \cite{Swan} for
$I=2$ $\pi\pi$ scattering
and was found to be small.

Weinberg \cite{Wein} used PCAC to predict an
$I=3/2$ $K\pi $ scattering length of
\begin{equation}
a_S^{(3)}  = -{m_K m_\pi \over m_K + m_\pi} \ {1 \over 8 \pi f^2}
\end{equation}
in his original PCAC paper.
Here, $f$ is the pseudoscalar decay constant which may be identified with
$f_\pi$ in the flavor symmetric limit.
The quark Born approximation for the
scattering length may be extracted from our expression for the S-wave phase
shift, and is

\begin{equation}
a_S^{(3)} = -{m_K m_\pi \over m_K + m_\pi}\;
{2 \alpha_s \over 9 m_u^2} \left[ 1 +
  \left( {4 \xi \over 1 + 2\xi}\right)^{3/2} +  \rho \left( {4 \over 2 +\xi}
\right)^{3/2} +  \rho \left( {2 \sqrt{\xi} \over 1 +\xi}\right)^3 \right] \ .
\end{equation}
With our various parameter sets we find the following values for
the scattering length $a_S^{(3)}$:
$-0.092/m_\pi$ ($\rho = 1, \xi = 1$); $-0.077/m_\pi$
($\rho = 0.677, \alpha_s = 0.6$); $-0.078/m_\pi$ ($\rho = 0.604,
\alpha_s = 0.634$); $-0.076/m_\pi$ (global fit). These are compared to the
PCAC prediction, one-loop chiral perturbation
theory and various model calculations in Table\ \ref{table1}.
Experimental values for the scattering length range from
$-0.07/m_\pi$  to $-0.14/m_\pi$, and are also summarized
in Table\ \ref{table1}. Note that
we may also interpret our scattering length as
a quark Born formula for $f_\pi $
if we assume the PCAC relation (7).
With the flavor-symmetric parameter set we find $f_\pi = 80$ MeV,
in reasonable agreement with the experimental value of 93 MeV.

Our theoretical values for the scattering length
are consistent with most
experimental results, but not with the most recent,
which is that of Estabrooks {\it et al}.
Lang and Porod \cite{LP} note that the
Estabrooks {\it et al.} S-wave phase shift agrees with previous data for
$m_{K\pi } \agt 1$ GeV, but is
somewhat larger in magnitude than previous measurements
for $m_{K\pi } \alt 1$ GeV. Presumably this leads to their rather
large scattering length. It would clearly
be useful to resolve this experimental
discrepancy, since only in this mass region is there any indication
of a possible disagreement between the S-wave phase shift and our predictions.
It would also be very useful to improve the accuracy of the P-wave
measurement, which is a sensitive test of flavor symmetry breaking,
and to extend the S-wave phase shift measurements to higher invariant masses
as a test of the extremum predicted and perhaps observed near 1.4 GeV.

\section{Conclusions}

We have calculated
$I = 3/2$ $K\pi $ elastic scattering phase shifts
using a Born-order constituent-exchange description
in the framework of
the nonrelativistic quark
potential model.

Extensive
previous work leads us to believe that
one gluon exchange combined with quark exchange
may accurately describe nonresonant
hadron scattering in certain channels including $I=3/2$ $K\pi$,
and that the Born
approximation to the scattering amplitude is an acceptable one. This
reaction
is appropriate
for testing this model of scattering because
t-channel pion exchange is forbidden by $G$-parity and the experimental phase
shift shows no evidence for s-channel resonance formation.

Since this model was previously found to describe
the related $I=2$ $\pi\pi$ S-wave phase shift accurately \cite{BS},
approximate flavor
symmetry leads us to expect
that the predicted $I = 3/2$ $K\pi $
S-wave phase shift should at least be
in qualitative agreement with the data. This is
indeed found to be the case. The agreement is considerably improved
when flavor symmetry is broken by assigning the strange quark
a mass consistent with standard quark model values.
The P-wave $K\pi $ phase shift however
is generated entirely by flavor symmetry breaking effects
(primarily by the strange-nonstrange quark mass difference in our model), and
is not present in $I=2$ $\pi\pi$ scattering.
The very reasonable result we find for the P-wave phase shift using
fitted S-wave parameters is therefore a nontrivial and successful test
of the model.
Finally, the model predicts an
S-wave scattering length of about $-0.077/m_\pi$, which is in the range of
reported experimental values and is commensurate with the predictions of chiral
perturbation theory.

Although we find a small negative D-wave phase shift as has been reported
experimentally, the magnitude and energy dependence are
not well reproduced. This, however, is a
small contribution to the scattering amplitude, and the various other
scattering mechanisms which have been neglected in this calculation
may be significant in this case, and should be investigated in future.

Weinstein and Isgur have also studied S-wave
$I=3/2$ $K\pi $ scattering in the nonrelativistic quark model, using a
nonperturbative variational technique \cite{WI2}.
(Their method does not allow extraction of higher partial waves
at present.)
They find good agreement
with the S-wave data, although they must first
scale the range and strength of their
effective $K\pi $ potentials.
We perform no such scaling
but do employ relativistic phase space; the fact that both methods agree
well with experiment suggests that their scaling may
actually be compensating for kinematic effects above threshold. This conclusion
is supported by a recent reanalysis of the Weinstein-Isgur variational
calculations \cite{JW}.

The obvious extension of this work is to meson-baryon and baryon-baryon
scattering, with the caveat that one should specialize to channels such as
K$^+$-nucleon and baryon-baryon in which $q\bar q$ pair creation and
annihilation is unimportant. These topics are currently under investigation.

\acknowledgements

This research was sponsored by the Natural Sciences and Engineering
Research Council of Canada; the United States Department of Energy under
contracts DE-AC05-840R21400 with Martin Marietta Energy Systems Inc.,
DE-FG05-91ER40627 with the Physics Department of the University of
Tennessee, DE-AC02-76ER03069 with the Center for Theoretical Physics at
the Massachusetts Institute of Technology; and by the State of
Tennessee Science Alliance Center under contract R01-1062-32.

\figure{S-wave $K\pi $ Phase Shift. The filled squares are data from
\cite{Esta}; the open squares are from \cite{Jong} (second solution). The
dotted
line corresponds to $\rho = 1$, the dashed line to $\rho = 0.677$, and the
solid line corresponds to the global fit (see text). \label{fig1}}

\figure{P-wave $K\pi $ Phase Shift. The data are from \cite{Esta}. The dashed
line corresponds to $\rho = 0.677$ and the solid line to the global fit (see
text). \label{fig2}}

\bigskip
\narrowtext
\begin{table}
\caption{Experimental and Theoretical Values for the S-wave Scattering Length.}
\begin{tabular}{lll}
$a_S^{(3)} \cdot m_\pi$ & Ref. & comments\\
\tableline
$-0.071(10)$ & \cite{Jong} & experimental \\
$-0.076(10)$ & \cite{ant} & experimental \\
$-0.084(11)$ & \cite{bak} & experimental \\
$-0.086(24)$ & \cite{kir} & experimental \\
$-0.091(9)$ & \cite{cho} & experimental \\
$-0.110(16)$ & \cite{ling} & experimental \\
$-0.138(7)$ & \cite{Esta} & experimental \\
       &            &           \\
$-0.05$ & \cite{BKM} & 1-loop chiral pert. theory \\
$-0.05$ & \cite{Pond} & pointlike meson model \\
$-0.06$ & \cite{Curry} & crossing-symmetric Regge model \\
$-0.07$ & \cite{Wein} & PCAC \\
$-0.074$ & \cite{lang} & coupled channel model \\
$-0.077$ & --- & this work (central value)\\
\end{tabular}
\label{table1}
\end{table}

\end{document}